\documentstyle[prd,aps,preprint,multicol]{revtex}

\begin{document}

\newcommand{\TeV}{\,{\rm TeV}}
\newcommand{\GeV}{\,{\rm GeV}}
\newcommand{\MeV}{\,{\rm MeV}}
\newcommand{\keV}{\,{\rm keV}}
\newcommand{\eV}{\,{\rm eV}}
\def\ap{\approx}
\def\beqar{\begin{eqnarray}}
\def\eeqar{\end{eqnarray}}
\newcommand{\bea}{\begin{eqnarray}}
\newcommand{\eea}{\end{eqnarray}}
\def\beq{\begin{equation}}
\def\eeq{\end{equation}}
\def\haf{\frac{1}{2}}
\def\plb#1#2#3#4{#1, Phys. Lett. {\bf #2B} (#4) #3}
\def\plbb#1#2#3#4{#1 Phys. Lett. {\bf #2B} (#4) #3}
\def\npb#1#2#3#4{#1, Nucl. Phys. {\bf B#2} (#4) #3}
\def\prd#1#2#3#4{#1, Phys. Rev. {\bf D#2} (#4) #3}
\def\prl#1#2#3#4{#1, Phys. Rev. Lett. {\bf #2} (#4) #3}
\def\mpl#1#2#3#4{#1, Mod. Phys. Lett. {\bf A#2} (#4) #3}
\def\rep#1#2#3#4{#1, Phys. Rep. {\bf #2} (#4) #3}
\def\lpp{\lambda''}
\def\ccg{\cal G}
\def\slash#1{#1\!\!\!\!\!/}
\def\rpv{\slash{R_p}}

\setcounter{page}{1}
\draft
\preprint{KAIST-TH 98/19, KIAS-P98040, hep-ph/9811363}

\title{Atmospheric and Solar Neutrino Masses
from Horizontal $U(1)$ Symmetry}

\author{Kiwoon Choi, Kyuwan Hwang}

\address{Korea Advanced Institute of Science and Technology,\\
        Taejeon 305-701, Korea}

\author{Eung Jin Chun}

\address{ Korea Institute for Advanced Study, \\
          207-43 Cheongryangri-dong, Dongdaemun-gu,
          Seoul 130-012, Korea}


\maketitle

\begin{abstract}
We study the neutrino mass matrix in supersymmetric models in which 
the quark and charged lepton mass hierarchies and also 
the suppression of baryon or lepton number violating couplings 
are all explained by  horizontal $U(1)_X$ symmetry.
It is found that the neutrino masses and mixing angles suggested by
recent atmospheric and solar neutrino experiments arise   
naturally in this framework which fits in best with
gauge-mediated supersymmetry breaking with large
$\tan\beta$.
This framework highly favors the small angle MSW oscillation of solar neutrinos,
and determine the order of magnitudes of all the neutrino mixing angles 
and mass hierarchies.
\end{abstract}



The fermion mass problem consists of understanding the flavor mixing
structure among quarks or leptons as well as the hierarchy of their
mass eigenvalues.  
It has been suggested that these hierarchical structures
can be explained by  a horizontal $U(1)_X$ symmetry 
whose spontaneous breaking is described by
$\lambda\ap$ Cabbibo angle \cite{FN,ETC,CL,CCK}.
Recent experimental data on atmospheric and solar neutrinos 
suggest non-vanishing neutrino masses and mixing \cite{SK-ATM,SOL}.
If spontaneously broken $U(1)_X$ is the origin of the quark
and charged lepton mass spectrum,
it is expected to have implications 
for the neutrino masses also.
It has been noted that when implemented in supersymmetric (SUSY) models,
$U(1)_X$ can explain not only the quark and lepton mass spectrum,
but also  the smallness of dangerous
baryon/lepton number ($B/L$) violating interactions \cite{CCK}.
This framework is interesting since renormalizable
$L$-violating couplings are small
enough to satisfy the current experimental bounds, but still nonvanishing
and thus can generate neutrino masses.
In this paper, we wish to examine the possibility that
the neutrino masses and mixing angles suggested by
recent atmospheric and solar neutrino experiments arise naturally  
in the framework of SUSY models in which 
the quark and charged lepton mass hierarchies and also 
the suppression of $B/L$-violating couplings 
are all explained by  horizontal $U(1)_X$ symmetry.
Combining the neutrino oscillation data 
with the informations from the quark and charged lepton sector and also 
the constraints on $B/L$-violating couplings, we find the $U(1)_X$ charge 
assignments producing all the fermion masses and mixing angles
correctly. 
This framework fits in best with gauge-mediated SUSY breaking
models with large $\tan\beta$,
favors the small angle MSW oscillation
of solar neutrinos over the large angle just-so oscillation,
and  determines the order of magnitudes of all
the neutrino mixing angles and mass eigenvalues.
In this framework, $m_2/m_3\ap 4\times 10^{-2}$ is essentially due to 
the loop to tree  mass ratio, while
$m_1/m_2\ap U_{e2}^2\ap \lambda^4$ is due to the $U(1)_X$ selection rule
where $m_A$ ($A=1,2,3$) denote the neutrino mass eigenvalues
and $U_{iA}$ the mixing matrix.


\medskip

To proceed, let us briefly summarize the  $U(1)_X$ selection rule
estimating the size of couplings \cite{FN,ETC,CL}.
The K\"{a}hler potential and superpotential of the model
are generically given by
\beqar
K &=& Z_{IJ}(\lambda,\bar{\lambda})
\Phi^I\Phi^{*J}+[X_{IJ}(\lambda,\bar{\lambda})\Phi^I\Phi^J+
{\rm h.c.}]+...,
\nonumber \\
W &=& \frac{1}{2}\tilde{\mu}_{IJ}(\lambda)\Phi^I\Phi^J+
\frac{1}{3 !}\tilde{Y}_{IJK}(\lambda)\Phi^I\Phi^J\Phi^K+...,
\eeqar
where $\Phi^I$ denote  light chiral superfields and
the ellipses stand for the terms of higher order in
$\Phi^I$.
The $U(1)_X$-breaking order parameter $\lambda$ 
corresponds to the VEV of a chiral superfield $\phi$
with the $U(1)_X$ charge $X(\phi)=-1$:
$\lambda=\langle \phi \rangle /M \ap 1/5$ for the fundamental
mass scale $M$ which is presumed to be of order the Planck scale
$M_P$.
The $U(1)_X$ selection rule
states that the hierarchical structures among the coefficients
are due to the insertion of $\lambda=\langle\phi\rangle/M$
or of $\bar{\lambda}=\langle\phi^*\rangle/M$
to make the corresponding operators to be $U(1)_X$-invariant.
This leads to
$Z_{IJ} \ap \lambda^{|x_I-x_J|}$,
$X_{IJ} \ap \lambda^{|x_I+x_J|}$ ,
$\tilde{\mu}_{IJ}\ap \tilde{\mu}\lambda^{x_I+x_J-X(\tilde{\mu})}$, 
$\tilde{Y}_{IJK}\ap \lambda^{x_I+x_J+x_K} \theta (x_I+x_J+x_K)$ 
where $x_I\equiv X(\Phi^I)$, i.e. the $U(1)_X$ charge of $\Phi^I$,
$\tilde{\mu}$ denotes the {\it representative} component
of $\tilde{\mu}_{IJ}$ whose operator has the $U(1)_X$ charge
$X(\tilde{\mu})$,
and $\theta(x)=1$ when $x$ is a non-negative integer,
while $\theta (x)=0$ otherwise.
The overall size of dimensionful $\tilde{\mu}_{IJ}$ 
depends upon the mechanism generating the corresponding bilinear terms
and can differ from the fundamental mass scale $M$ in general.

After integrating out supersymmetry
breaking fields while taking into account supergravity effects,
one can redefine the chiral superfields in the resulting
effective theory
to have a  {\it canonical} K\"{a}hler metric: $\Phi^I\rightarrow R_{IJ}\Phi^J$
where $R_{IJ}$ obeys
$R_{IJ}Z_{JK}R^*_{KL}=\delta_{IL}$.
The order of magnitude estimate of $Z_{IJ}$ above  implies
$R_{IJ}\ap \lambda^{|x_I-x_J|}$.
Then for the redefined $\Phi^I$ with {\it canonical} kinetic term, 
the bilinear and trilinear couplings of 
the  effective superpotential are given by
\beqar
&& \mu_{IJ} \ap \tilde{\mu}_{IJ}+R_{IK}R_{JL}\tilde{\mu}_{KL}
+m_{3/2}X_{IJ}
\nonumber \\
&&
Y_{IJK} \ap \tilde{Y}_{IJK}+R_{IL}R_{JM}R_{KN}\tilde{Y}_{LMN} ,
\eeqar
including first  the contribution from the bare superpotential
$W$,
second the effects of  superfield redefinition $\Phi^I\rightarrow
R_{IJ}\Phi^J$,
and finally the supergravity contribution  from
the Kahler potential which is proportional to the gravitino mass
$m_{3/2}$.

The most general $SU(3)_c\times SU(2)_L\times
U(1)_Y$-invariant superpotential of the MSSM superfields
is given by
\beqar 
  W_{\rm MSSM} &&= \mu H_1 H_2 + Y^u_{ij} H_2 Q_i U^c_j +
         Y^d_{ij} H_1 Q_i D^c_j + Y^e_{ij} H_1 L_i E^c_j \nonumber\\
  &&+ \Lambda^u_{ijk}  U^c_i D^c_j D^c_k +
         \Lambda^d_{ijk} L_i Q_j D^c_k + \Lambda^e_{ijk} L_i L_j E^c_k
\nonumber \\
&&+\frac{1}{M_S}\Gamma_{ij}L_iH_2L_jH_2+ ..., \label{mssm}
\eeqar
where $(H_1,H_2)$, $(L_i, E^c_i)$, $(Q_i, U^c_i, D^c_i)$  denote the Higgs,
lepton, and quark superfields, respectively.
Among the possible non-renormalizable
operators,  we include only the $d=5$ see-saw operator which is presumed
to be induced by the physics at the scale $M_S$.
Here we have rotated away the possible bilinear term $\mu_iL_iH_2$
through the unitary rotation of superfields, which does not alter
the order of magnitude estimates of couplings.

The $U(1)_X$ charges denoted by the small letters
$q_i, u_i,$ e.t.c. for the  superfields 
$Q_i, U^c_i,$ e.t.c. are well constrained by the experimental data.
The quark Yukawa couplings $Y^{u,d}_{ij}$ 
are determined by the $U(1)_X$ charges of the operators $H_2 Q_i U^c_j$ and
$H_1 Q_i D^c_j$.  
The large $m_t\ap \langle H_2\rangle$ suggests
first of all $q_3+u_3+h_2=0$,  while
$m_b \tan\beta \ap \lambda^x m_t$ where $x=q_3+d_3+h_1$ and
$\tan \beta=\langle H_2\rangle/\langle H_1\rangle$.
Then the observed pattern of the CKM matrix and the quark mass eigenvalues
 $(m^u)\ap  m_t (\lambda^8, \lambda^4, 1)$ and 
 $(m^d)\ap m_b (\lambda^4, \lambda^2, 1)$
contain enough informations 
to reconstruct the  Yukawa matrices $Y^u$ and $Y^d$, 
leading to \cite{CL}
$$ (q_{13},q_{23}, u_{13},u_{23}, d_{13},d_{23})$$
\beq
= (3,2,5,2,1,0) \quad {\rm or} \quad 
(-3,2, 11,2,7,0),
\eeq
where $q_{ij}=q_i-q_j$, e.t.c.,
and $x<3$ for the second case.
The charged lepton mass hierarchy  
$(m^e) \ap  m_\tau (\lambda^5, \lambda^2, 1)$ suggests
$(e_{13},e_{23}) = 
	    (5-l_{13},2-l_{23})$ 
(Case I)
or $(9-l_{13},-2-l_{23})$ (Case II) where
$e_{ij}=e_i-e_j$, e.t.c.
In order to have $Y^{e}_{12,21}\lesssim \lambda^{2+x}$ and also a
non-singular $Y^e_{ij}$, we need more constraints:
$l_{13} \ge l_{23}$ or $l_{13} < l_{23} - (x+2)$
for Case I,
$-2 \leq l_{23}\leq 0$,  $x<2$, 
$l_{13} \geq l_{23}+4$ for $l_{23}>x-2$
for Case II.
As we will see, the neutrino oscillation data
imply $l_{23}=0$ and $l_{13}=2$, which is compatible only with
Case I. We thus have
\beq
(e_{13},e_{23}) = 
	    (5-l_{13},2-l_{23})=(3,2).
\eeq

The lepton flavor mixing is usually read off from 
the non-trivial neutrino mass matrix $m^{\nu}_{ij}$
in the basis of the diagonal charged 
lepton mass matrix.  Denoting the unitary mixing matrix by $U_{iA}$,
the mass eigenvalues are given by
$m_A=U_{iA}U_{jA} m^{\nu}_{ij}$ ($A=1,2,3$).
Recent Super-Kamiokande and other experiments
on the atmospheric and solar neutrino oscillations \cite{SK-ATM,SOL}
suggest the following neutrino oscillation parameters:
\begin{eqnarray} \label{ATM} 
&&~~~\Delta m^2_{32} \approx \Delta m^2_{31}\approx 
 2.2\times10^{-3}{\rm eV}^2\,, \quad \theta_{\rm atm} \approx 1 \,,
\nonumber \\
&&\left\{ \begin{array}{lcl}
 \Delta m^2_{21} &\approx&  5\times 10^{-6} {\rm eV}^2\,, 
          \quad \theta_{\rm sol} \approx 3.7\times10^{-2}
\quad \leftarrow \mbox{MSW} \\
\Delta m^2_{21} &\approx&  6.5\times 10^{-11} {\rm eV}^2\,,
          \quad \theta_{\rm sol} \approx 0.52
\quad  \leftarrow \mbox{just-so}
\end{array}\right.   \label{sold}
\end{eqnarray} 
where $\Delta m^2_{AB}=m^2_A-m^2_B$,
$\theta_{\rm atm}=\theta^{32}_{\mu\tau}$
or $\theta^{31}_{\mu\tau}$, and 
$\theta_{\rm sol}=\theta^{21}_{e\mu}$ or $\theta^{21}_{e\tau}$
for the effective mixing angle:
$\sin^22\theta^{ij}_{\alpha\beta}=4 
|U_{\alpha i} U^*_{\alpha j} U^*_{\beta i} U_{\beta j}|$.  

\medskip

In our framework, 
$L$-violating couplings are suppressed 
by having $l_i$ significantly bigger than $h_1$ and $-h_2$.
As will be shown explicitly later,
the covariance under $U(1)_X$ suggests that
the neutrino mass matrix in our framework takes the form:
\beq
(m^{\nu})_{ij}=m_3 \lambda^{l_{i3}+l_{j3}} A_{ij},
\label{mmat}
\eeq
where $m_3$ is the largest mass eigenvalue,
 $l_{i3}=l_i-l_3$, and all $A_{ij}$ are of order unity.
This form of $m^{\nu}$ leads first of all to 
$U_{i 3}\ap \lambda^{l_{i 3}}$.
The large atmospheric $\nu_{\mu}$-$\nu_{\tau}$ mixing unambiguously implies
$U_{\mu 3}\ap 1$,  and thus 
\beq
l_2=l_3.
\eeq
For $l_2=l_3$, the mass matrix (\ref{mmat}) implies also 
$U_{i 2}\ap \lambda^{l_{i3}}$. Combined with the unitarity,
this determines the mixing matrix to take the form:
\begin{equation} \label{}
U \ap \left( \begin{array}{ccc} 1 & \lambda^{l_{13}} & \lambda^{l_{13}} \\
                                       \lambda^{l_{13}} & 1 & 1 \\
                                       \lambda^{l_{13}} & 1 & 1 
                     \end{array} \right) \, . 
\end{equation}
Given the above form of $U$, the MSW mixing angle 
$\theta_{\rm sol} \approx \lambda^2$ implies
\beq
  l_1=l_2+2=l_3+2 , \label{l13}
\eeq
while the just-so oscillation leads to
$l_1=l_2=l_3$.
Recent Super-Kamiokande and CHOOZ data \cite{SK-ATM,CHOOZ}
indicates that $\nu_\mu$ rarely if ever oscillates into $\nu_e$,
which can be interpreted as excluding $l_1=l_2=l_3$ \cite{GG}.


We have seen that 
all $U_{iA}$ resulting from the mass matrix (\ref{mmat})
are determined essentially by the $U(1)_X$ charges $l_i$.
As it will become clear later, although all $A_{ij}$ are of order unity,
the corresponding matrix is naturally {\it approximately singular}
and thus gives a mass hierarchy $m_3\gg m_2\gg m_1$.
Then the oscillation data (\ref{ATM}) implies 
\beq
m_3\approx 5\times 10^{-2} \, {\rm eV},
\quad m_2/m_3\ap 4\times 10^{-2}.
\eeq

Let us now discuss how the neutrino mass matrix (\ref{mmat}) with 
approximately singular $A_{ij}$ arises 
in SUSY models with $U(1)_X$.
Although not a unique possibility, an attractive
scheme to suppress dangerous $L$-violating couplings in our framework is 
to have $l_i$ significantly bigger than $h_1$ and $-h_2$.
In this scheme, 
one can easily arrange the physics at $M_S$,
e.g. the $U(1)_X$ charges of the superheavy singlet neutrinos,
to make  the resulting see-saw coefficients $\Gamma_{ij}$ 
in (\ref{mssm}) suppressed by 
$\lambda^{l_i+l_j+2h_2}$ \cite{grossman}.
If $M_S$ is the string scale
$M_{\rm string}\ap 5\times 10^{17}$ GeV or
the unification scale $M_{\rm GUT}\ap 2\times 10^{16}$ GeV, which is 
perhaps the most plausible possibility,
this would result in
\beq
m^{\rm seesaw}_{ij} \ap 
(10^{-3}\sim 10^{-4}) \times 
\lambda^{l_i+l_j+2h_2}
\,  {\rm eV},
\eeq
which is too small to be relevant 
for the atmospheric  and solar neutrino masses for
$l_i$ significantly bigger than $-h_2$.
In fact, the two representative models that we found in this paper
have $l_i+h_2\geq 8$ and thus a completely negligible see-saw contribution.

Once the see-saw contribution is negligibly small,
the atmospheric and solar neutrino masses arise from
the renormalizable  interactions in 
the superpotential (\ref{mssm}) and also the following soft SUSY breaking 
terms \cite{HEM,CKKL,JV}: 
\bea
&V_{\rm soft}&= m^2_{L_iH_1}L_iH_1^*+ B_iL_iH_2 
+ A^{d}_{ij} H_1 Q_i D^c_j \nonumber \\
&+& A^{e}_{ij} H_1 L_i E^c_j
       + C^{d}_{ijk} L_i Q_j D^c_k + C^{e}_{ijk} L_i L_j E^c_k
       \,\, + {\rm h.c.}
\eea
where now all field variables denote the scalar components of the 
corresponding superfields.
The $L$-violating $B_i$ or
$m^2_{L_iH_1}$ (in the basis where $\mu_iL_iH_2$ in the
superpotential are rotated away)
results in the tree-level neutrino mass \cite{HS,TREE}:
\begin{equation} \label{mntree}
 m^{\rm tree}_{ij} \ap
  \frac{g_a^2 \langle\tilde{\nu}^*_i\rangle \langle\tilde{\nu}^*_j\rangle} 
{M_a} \,,
\end{equation}
where $M_a$ denote the $SU(2)\times U(1)$  gaugino masses and the sneutrino VEV's are given by  
\beq
\langle\tilde{\nu}^*_i\rangle \,\,\ap\, \frac{2M_Z
(m^2_{L_iH_1}\cos\beta +B_i\sin\beta)}{
        m^2_{\tilde l}+\frac{1}{2}M_Z^2\cos 2\beta},
\eeq 
for the $Z$-boson mass $M_Z$ and
$m_{\tilde{l}}$ denoting the slepton soft mass
which is assumed to be (approximately) flavor-independent.
There are also the contribution from the finite 1-loop
graph involving squark or slepton exchange in the $\mu_{i}=0$ basis:
\bea   \label{mnloop}
m^{\rm loop}_{ij} &=& {1 \over 16\pi^2}   \left[
  3 \, \frac{\Lambda^d_{ilm}\Lambda^d_{jnk}
        Y^{d*}_{lk}\langle H^{*}_1\rangle
  \left(A^{d*}_{nm}\langle H^{*}_1\rangle
        + \mu Y^{d*}_{nm} \langle H^{*}_2\rangle\right)
}{m^2_{\tilde q}} \right. \\
&+&\left.
  \frac{\Lambda^e_{i\alpha m}\Lambda^e_{j\beta k}
        \Lambda^{e*}_{\gamma\alpha k}\langle L^{*}_{\gamma}\rangle
  \left(C^{e*}_{\delta\beta m}\langle L^{*}_\delta\rangle
        + \mu\Lambda^{e*}_{0\beta m}
        \langle H^{*}_2\rangle\right)
}{m^2_{\tilde l}}+(i \leftrightarrow j) \right]\,,
\nonumber
\eea
where the Greek indices 
($\alpha, \beta$, ...) run from 0 to 3 with $L_0\equiv
H_1$, while the Roman indices ($i,j$,...) run from 1 to 3.
Here $\Lambda^e_{i0k} = -\Lambda^e_{0jk} \equiv Y^e_{ik}$,
$C^e_{i0k} = -C^e_{0jk} \equiv A^e_{ik}$, and
${m}^2_{\tilde{q}}$ and ${m}^2_{\tilde{l}}$ denote 
the squark and slepton soft masses 
which are assumed to be (approximately) flavor-independent.
Note that all the parameters in (\ref{mntree}) and (\ref{mnloop}) are
renormalized at the weak scale.
The contribution involving the sneutrino VEV
$\langle L_i\rangle\equiv\langle\tilde{\nu}_i\rangle$
in the loop mass has been overlooked so far,
but turns out to be crucial to fit $m_2/m_3\ap 4\times 10^{-2}$
in our framework.


\medskip

If $U(1)_X$ is anomalous, which is the most interesting possibility, 
the quadratically divergent Fayet-Iliopoulos  coefficient 
$\lambda^2 M_P^2$ naturally yields $\langle\phi\rangle/M_P\ap \lambda$.
It also leads to a nonvanishing $U(1)_X$ $D$-term
\cite{dvali,casas}:
\beq
D_X\ap |F|^2/M_X^2
\ap |F|^2/g^2\lambda^2 M_P^2,
\eeq
where $F$ denotes the  SUSY-breaking $F$-term, $M_X\ap g \langle \phi\rangle$ 
the $U(1)_X$ gauge boson mass.  
The soft scalar mass of $\Phi^I$ then receives
a $D$-term contribution $\delta m_I^2= X(\Phi^I) D_X$.
In gravity-mediated SUSY-breaking models, this $D$-term contribution
dominates over the standard $F$-term contribution $|F|^2/M_P^2$.
If the $U(1)_X$ charge  $X(\Phi^I)$ is {\it flavor-independent}, 
the scalar masses dominated by the $D$-term contribution
would be (approximately)  degenerate, thereby
avoid the dangerous flavor violation \cite{dvali,moha}.
However in our framework, $X(\Phi^I)$  are {\it flavor-dependent}
to explain the fermion mass hierarchy.
When the $D$-term contribution is important,
the requirement to avoid dangerous flavor violation 
while  explaining the quark and charged lepton mass spectrum
through {\it flavor-dependent} $X(\Phi^I)$
severely constrains the possible $U(1)_X$ charge assignment \cite{nelson}, 
and actually
leads to the so-called ``more" minimal supersymmetry \cite{kaplan}.
However the resulting $X(\Phi^I)$  do {\it not} fit in with
our framework explaining the small $B/L$-violating couplings
by means of $U(1)_X$.
It thus appears that gravity-mediated models with $U(1)_X$ do {\it not}
fit in  well with our framework.

The bothersome {\it flavor-dependent} $U(1)_X$ $D$-term contribution  
becomes negligible 
in gauge-mediated SUSY-breaking models with a messenger scale
$M_m\ll \frac{\alpha}{4\pi} M_X$ 
for which $\sqrt{D_X}\ll m_{\rm soft}\ap \frac{\alpha}{4\pi}F/M_m$.
To discuss the neutrino mass matrix in gauge-mediated case,
let $\mu_{\alpha}L_{\alpha}H_2$
and $B_{\alpha}L_{\alpha}H_2$
denote the bilinear terms in
the superpotential and soft scalar potential in generic basis,
and $m^2_{\alpha\beta}L_{\alpha}L^*_{\beta}$ the soft scalar masses of
$L_{\alpha}=(H_1,L_i)$.
If the messenger gauge interactions
do {\it not} distinguish $H_1$ from $L_i$,
$B_{\alpha}$ is naturally aligned to $\mu_{\alpha}$ and also
$m^2_{\alpha\beta}=m_0^2\delta_{\alpha\beta}$ at 
$M_m$.
In this case, $B_i$ and $m^2_{L_iH_1}$ can be simultaneously rotated 
away as $\mu_i$ at $M_m$, i.e. $B_i(M_m)=m^2_{L_iH_1}(M_m)=0$ in the basis
of $\mu_i=0$,
and their low energy values at $M_Z$ are determined 
by the RG evolution which is governed by
the $\Delta L=1$ Yukawa couplings $\Lambda^{d,e}_{ijk}$ and 
generic $L$-conserving couplings. 
The $L$-violating trilinear soft scalar couplings ($C^{d,e}_{ijk}$)
at $M_Z$ are also determined by the RG evolution.
Thus in gauge-mediated models,
all renormalizable $L$-violating couplings at $M_Z$ are calculable
in terms of $\Lambda^{d,e}_{ijk}$  and also of generic $L$-conserving
couplings. 

\medskip

Soft parameters  in  gauge-mediated models \cite{gauge} typically satisfy:
$M_a/\alpha_a\ap m_{\tilde q}/\alpha_3\ap
m_{\tilde l}/\alpha_{1,2}$ at the gauge messenger scale $M_m$
where $M_a$, $m_{\tilde q}$, and $m_{\tilde l}$  denote the gaugino,
squark and slepton masses, respectively,
and $\alpha_a=g_a^2/4\pi$
for the (SU(5)-normalized) standard model gauge coupling constants 
$g_a$ ($a=1,2,3$).
Trilinear scalar coefficients do vanish at $M_m$
and thus their low energy values are determined by the RG evolution.
The size of the bilinear term $BH_1H_2$ in the scalar potential
depends upon how $\mu$ is generated. 
An attractive possibility in this regard 
is $B(M_m)=0$ for which
all CP-violating phases in soft parameters at $M_Z$ are automatically
small enough to avoid a too large electric dipole moment \cite{dine,RS}.
In this case,
the RG-induced low energy value of $B$ yields
a large $\tan\beta\ap (m_{H_1}^2+m_{H_2}^2+2\mu^2)/B(M_Z)=40\sim 60$
which corresponds to $x=0$ in view of $\tan\beta\ap
\lambda^x m_t/m_b$.
In fact, a careful analysis of the neutrino mass matrix
implies that when $x\geq 1$ it is rather difficult to fit 
$m_2/m_3\ap 4\times 10^{-2}$
for reasonable range of soft parameters in gauge-mediated models \cite{JV}
without a sizable cancellation
\cite{cch}, and thus here we concentrate on $x=0$.

Analyzing the neutrino masses (\ref{mntree}) and (\ref{mnloop}) determined
by the RG evolution of couplings with the boundary conditions that
trilinear soft scalar couplings, $B$,
$B_i$ and $m^2_{L_iH_1}$ are all vanishing  at $M_m$,
and also $M_a/\alpha_a\ap m_{\tilde{q}}/\alpha_3\ap m_{\tilde{l}}/\alpha_{1,2}$
at $M_m$,
it is straightforward to find that (for $x=0$
and thus $\tan\beta=40\sim 60$)
\beq
m^{\rm tree}_{ij}= 
10^{-1}\xi_1 t^4 a_ia_j 
\left(\mu^2M^2_Z \over m^3_{\tilde{l}}\right) \,,
\label{numass1}
\eeq
where $a_i\ap Y_b\Lambda^d_{i33}
\ap \lambda^{l_i-h_1}$, $t=
\ln (M_m/m_{\tilde l})/\ln (10^3)$
 and $\xi_1$ is the coefficient of order
unity summarizing the uncertainty of our estimate.
Among various terms in the loop mass (\ref{mnloop}),
the leading contribution to the loop mass comes from the piece involving 
$\langle L_i\rangle\langle H_2\rangle$ for large $\tan\beta$,
because  $Y^e \langle L_i \rangle/\Lambda^e \langle H_1 \rangle 
\ap \tan\beta\gg 1$. 
We then have
\beq
m^{\rm loop}_{ij}\ap 10^{-2}\xi_2 t^2
Y_bY_{\tau}^3\Lambda^d_{333}(\delta_{i3}\Lambda^e_{j33}
+\delta_{j3}\Lambda^e_{i33}) 
\left( \mu^2 M_Z^2 \over m^3_{\tilde{l}}\right)  \,,
\label{numass2}
\eeq
where the smaller contributions are  ignored
and again the coefficient  $\xi_2$ of order unity is introduced to
take into account the uncertainty of our estimate.
Here the powers of $t\propto\ln (M_m/m_{\tilde l})$ are from
$\langle \tilde{\nu}_i\rangle\propto B_i(m_{\tilde l})
\propto t^2$ under the boundary condition
$B_i(M_m)=0$.
The above neutrino mass matrices are derived in the basis for which
$Y^e_{ij}$ and $Y^d_{ij}$ are diagonal, and $Y_b=Y^d_{33}$
and $Y_{\tau}=Y^e_{33}$.
At any rate, from (\ref{numass1}) and (\ref{numass2}), we find
\beq
\frac{m^{\rm loop}_{23}}{m^{\rm tree}_{23}}
\ap 10^{-2}
\left(\frac{\xi_2\Lambda^e_{233}}{\xi_1\Lambda^d_{233}}\right)
\left(\frac{\tan\beta}{50}\right)^2
\left( \ln10^3 \over \ln{M_m \over m_{\tilde{l}}}\right)^2,
\eeq
where we have used $Y_{\tau}\ap \tan\beta/95$, $Y_b\ap \tan\beta/50$.
Since $\Lambda^e_{233}$ and $\Lambda^d_{233}$ are comparable
to each other, the above result 
shows that $m^{\rm tree}_{ij}$ gives a dominant contribution.

Obviously $m^{\rm tree}_{ij}$ is a rank 1 matrix, and thus
the total neutrino mass matrix takes the form (\ref{mmat})
with an approximately singular matrix $A_{ij}$ when $m^{\rm tree}_{ij}$
dominates.
We then find from (\ref{numass1}) and (\ref{numass2})
the following mass hierarchies:
\bea
&& m_3\ap U_{i3}m^{\rm tree}_{ij}U_{j3}\ap  10^{-1}
\eta M_Z 
\lambda^{2(l_3-h_1)} \,, 
\nonumber \\
&& m_2\ap U_{i2}m^{\rm loop}_{ij}U_{j2}\ap 
m_3 {m^{\rm loop}_{23} \over m^{\rm tree}_{233}}\,,
\nonumber \\
&& m_1\ap U_{i1}m^{\rm loop}_{ij}U_{j1}\ap
m_2 \lambda^4\,, 
\eea
where $\eta=\xi_1
(\ln{M_m\over m_{\tilde{l}}}/ \ln 10^3)^4
(M_Z\mu^2 / m_{\tilde l}^3)$.
For $m_{\tilde l}\ap 200 \sim 400 \GeV$ and 
$\mu\ap 2m_{\tilde l}$ 
which has been suggested to be the best parameter range for 
correct electroweak symmetry breaking \cite{RS}, $\eta$
is roughly of order unity and then
the experimentally favored  $m_3\ap 5\times 10^{-2}$ eV 
can be obtained for
\begin{equation}
7\lesssim  l_3 - h_1 \lesssim 9. 
\end{equation}
Note that in our framework small $m_2/m_3$ is essentially due to
the loop to tree mass ratio, while the other small mass ratios
$m_1/m_2\ap \lambda^4$ and $m_3/M_Z\ap 10^{-1}\lambda^{2(l_3-h_1)}$ 
are from the $U(1)_X$ selection rule.

We found many possible $U(1)_X$ charge assignments producing all fermion
masses and mixing discussed so far, while satisfying all the bounds
on $B/L$-violating couplings \cite{RR}
through the $U(1)_X$ selection rule under the condition 
that the maximum $U(1)_X$ charge is not unreasonably large
for $X(\lambda)=-1$.
For more detailed discussions,
see \cite{cch}.
In this paper, we  pick two representative solutions: Model 1 and Model 2
which are listed in Table 1.

\medskip

To conclude, we have studied  
the neutrino mass matrix 
in supersymmetric models in which 
the observed quark and charged lepton 
masses and also the suppression of $B/L$ violating couplings are all
explained by horizontal $U(1)_X$ symmetry.
A particular attention was paid for
the possibility that the neutrino masses and mixing angles
suggested by recent atmospheric and solar neutrino experiments 
arise naturally in this framework.
It is found that our framework
fits in best with gauge-mediated SUSY breaking
models with large $\tan\beta\ap 50$, and 
favors the small angle MSW oscillation
of solar neutrinos over the large angle just-so oscillation.
Combining the informations from neutrino oscillation experiments
with those from the quark and charged lepton sector and also the constraints
on $B/L$-violating couplings, we find the $U(1)_X$ charge 
assignments producing all the fermion masses and mixing angles
correctly. This framework determines
the order of magnitudes of the neutrino mixing matrix elements and  mass eigenvalues
to be:
$U\ap U^T$ with $U_{e2}\ap U_{e3} \ap \lambda^{l_3-l_1}=\lambda^2$,
$ U_{\mu 3}\ap \lambda^{l_3-l_2}\ap 1$ and  
$m_1/m_2\ap \lambda^{2(l_3-l_1)}\ap \lambda^4$, 
$m_2/m_3\ap ({\rm Loop}/{\rm Tree})\ap {\cal O}(10^{-2})$ 
for $m_3\ap {\cal O}(10^{-1})\times  M_Z \lambda^{2(l_3-h_1)}\ap 5\times 10^{-2}$ eV.

{\bf Acknowledgments}:
KH would like to thanks H. D. Kim for helpful discussions.
This work is supported in part
by KOSEF Grant 981-0201-004-2,
MOE Basic Science Institute Program BSRI-98-2434,
KOSEF through CTP of Seoul National University,
and KRF under the Distinguished Scholar Exchange Program.

\def\hh{@{\hspace{5.9mm}}}

\renewcommand{\arraystretch}{1}
\begin{table}
\caption {
$U(1)_X$ charge assignments for the MSSM fields of two representative
models.
}
\bigskip
\begin{tabular}{|c|c|c|c|c|c|c|}
Model & $q_1,q_2,q_3$ & $ u_1,u_2,u_3$ &
$d_1,d_2,d_3$ & $l_1,l_2,l_3$ &
$e_1,e_2,e_3$ & $h_1,h_2$  \\ \hline
 1     & 8,7,5 &-3,-6,-8 & -1,-2,-2 & 7,5,5 & 1,0,-2 & -3,3 \\ \hline
 2     & 7/2,5/2,1/2 & -1/2,-7/2,-11/2 & 11/2,9/2,9/2 & 5,3,3 & 5,4,2 & -5,5 
 \\ 
\end{tabular}
\vspace{1cm}
\end{table}

\end{document}